\documentclass[twocolumn,english,journal]{IEEEtran}
\pdfoutput=1
\usepackage[T1]{fontenc}
\usepackage[latin9]{inputenc}
\usepackage[active]{srcltx}
\usepackage{color}
\usepackage{cite}
\usepackage{babel}
\usepackage{array}
\usepackage{prettyref}
\usepackage{amsmath}
\usepackage{amssymb}
\usepackage{graphicx}
\usepackage{epstopdf}
\usepackage{url}
\usepackage{enumitem}
\usepackage[short]{optidef}
\usepackage{bm}
\usepackage{smartdiagram}
\usesmartdiagramlibrary{additions}
\graphicspath{{figures/}{photos/}}

\makeatletter

\usepackage[caption=false,font=footnotesize]{subfig}
\newcommand*{\tdnetgen}{\emph{TDNetGen}}

\makeatother

\begin{document}

\title{TDNetGen: An open-source, parametrizable, large-scale, transmission and distribution test system}

\author{Nicolas~Pilatte,~Petros~Aristidou,~\IEEEmembership{Member,~IEEE},
and~Gabriela~Hug,~\IEEEmembership{Member,~IEEE}%
\thanks{Nicolas~Pilatte and Gabriela~Hug are with the Power Systems Laboratory, ETH Zurich, Physikstrasse 3, 8092 Zurich, Switzerland (e-mail: \mbox{npilatte@student.ethz.ch}; \mbox{hug@eeh.ee.ethz.ch})}%
\thanks{Petros~Aristidou is with the School of Electronic and Electrical Engineering, University of Leeds, Leeds LS2 9JT, UK (e-mail: \mbox{p.aristidou@ieee.org})}%
}
\maketitle

\begin{abstract}
In this paper, an open-source MATLAB toolbox is presented that is able to generate synthetic, combined transmission and distribution network models. These can be used to analyse the interactions between transmission and multiple distribution systems, such as the provision of ancillary services by active distribution grids, the co-optimization of planning and operation, the development of emergency control and protection schemes spanning over different voltage levels, the analysis of combined market aspects, etc. The generated test-system models are highly customizable, providing the user with the flexibility to easily choose the desired characteristics, such as the level of renewable energy penetration, the size of the final system, etc.
\end{abstract}


\section{Introduction}

\IEEEPARstart{T}{he} most noticeable developments foreseen in the near future in power systems involve Distribution Networks (DNs). Future DNs are expected to host a big percentage of Renewable Energy Sources (RES) and other Distributed Energy Resources (electric vehicles, flexible loads, fuel cells, batteries, etc.). Moreover, it is expected that DNs will be called upon to actively support the bulk Transmission Network (TN) participating in ancillary services with the help of Information and Communication Technologies (ICT) and advanced management and control techniques. For these reasons, the interaction between the Transmission and Distribution (T\&D) grids has become the focus in many research areas of power systems over the last years. It has drawn significant attention in the areas of voltage stability and support~\cite{Aristidou2015d, Lin2016, Ding2017}, combined system dynamic stability~\cite{TR22}, optimisation of power and reserves~\cite{Caramanis2016,Li2016}, dynamic simulations for security assessment \cite{Li2015,Aristidou2015}, and much more. This increased research interest has been coupled with an equally high number of national and international research projects, funded both by industry and governmental agencies, to define or analyse the interactions between Transmission System Operators (TSOs) and Distribution System Operators (DSOs), e.g.~\cite{TDI2,SmartNet,evolvdso}.

\subsection{Review of popular test system models}

Several test systems exist for separately studying TNs or DNs. For TN studies, the most widely used systems are the ones developed by IEEE, for power flow~\cite{dataieee, dataieee2} and transient stability studies~\cite{Ieee1992,Demetriou2015}. More specialised test system have also been proposed, for instance, the RTS-79 and RTS-96~\cite{rts96} for reliability studies, or the revised Nordic system for dynamic and voltage stability studies~\cite{PES-TR19}. Several variants of these models have been published over the years depending on the type of study and the phenomena tackled by the method being tested. Other research groups and organisations have also developed their own test systems; for example, the 150-bus synthetic system from the University of Illinois, Urbana-Champaign~\cite{uiuc150}, the 9-bus and 179-bus systems from the Western System Coordinating Council~\cite{wscc9}, used for transient stability studies~\cite{wscc9_bis,Singh2015}, and more~\cite{IEEJ,Semerow2015,CigreTFC6.04.02/TB5752014,Josz2016}.

Regarding distribution network models, a wide variety of distribution test systems have been proposed by IEEE and can be found in \cite{dn_systems}. In addition, the 33-bus and the 69-bus test systems \cite{baran1989network,baran1989optimal} have been used, among other things, to study the impact of distributed generation~\cite{Acharya2006} while the RBTS 6-bus test system has been used for reliability analysis~\cite{Wang1993}. Several other, isolated or collective, efforts have been made to prepare and make available generic DN test systems, e.g.~\cite{UKDGS,CigreTFC6.04.02/TB5752014}.

While there is an abundance of individual TN and DN systems, there is a lack of combined T\&D systems allowing to investigate the interplay between them. Such systems are required to examine the impact of active distribution networks on the TN, study various proposed methods for DNs to support the TN in steady-state or during fault dynamics, develop or validate DN equivalent models that can later be used in bulk TN stability studies, test T\&D co-optimization algorithms (for planning or operation), analyze the techniques for including DNs to power markets, and many more. 

In the past, specific test systems, such as in~\cite{Aristidou2015d}, have been developed by individual researchers or groups to analyse the T\&D interactions. However, constructing such a system is a tedious task with many challenges and parameters to be selected. In addition, the data of these systems is rarely published and the customization performed to match the specific problem studied, makes them difficult to be used in other applications. 

\subsection{Contributions}

In this paper, an open-source MATLAB toolbox named \tdnetgen\cite{TDNETGEN} is presented, that is able to generate large-scale T\&D network models (high and medium voltage) that can be used for a variety of studies. The generated model data can be freely modified and shared, allowing researchers to compare the performance of their algorithms against each other. \tdnetgen~allows to select several key characteristics of the generated system, such as the RES penetration, the scaling factors for loads and generators, etc. (some of these are detailed in \prettyref{sec:key_parameters}). 

Based on these characteristics, \tdnetgen~generates the combined T\&D test system using the well-known Nordic TN model, detailed in~\cite{PES-TR19}, and systematically replacing the aggregated TN loads with a detailed DN model, derived from~\cite{UKDGS} and customised to accommodate increased distributed generators (DGs) penetration. The toolbox is powered by MATPOWER~\cite{Zimmerman2011}, a well-known, open-source, steady-state, planning and analysis tool. The generated models can be exported in the native MATPOWER format or custom exporters can be easily implemented to allow importing to other software.

\begin{figure}
    \centering
    \includegraphics[width=\columnwidth]{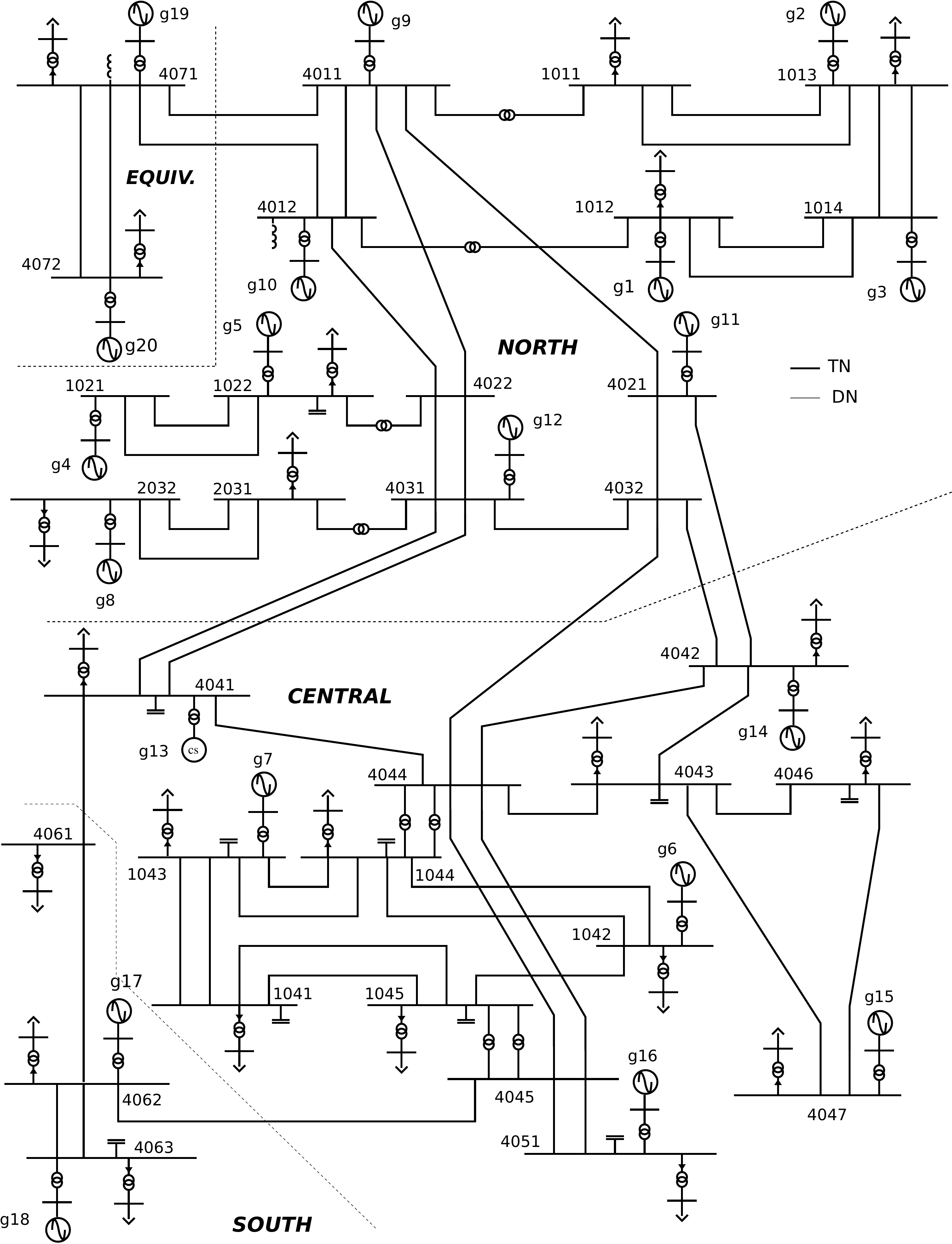}\caption{One-line diagram of the transmission network \cite{PES-TR19}}\label{fig:one_line_tn}
\end{figure}

\subsection{Paper Structure}

The remaining paper is organised as follows. In \prettyref{sec:systems}, the "template" TN and DN models that are used to generate the combined T\&D system are presented. Then, in \prettyref{sec:key_parameters}, the key parameters for selecting the characteristics of the generated system are explained and in \prettyref{sec:Methodology}, the procedure of generating the combined T\&D system is detailed. Finally, some example test cases are given in \prettyref{sec:Power-Flow-Test-Cases} followed by some concluding remarks in \prettyref{sec:conclusion}.

\section{Template test systems}\label{sec:systems}

In this section, we briefly present the TN and DN models that are combined to generate the resulting T\&D system.

The TN is based on the model documented in \cite{PES-TR19} and sketched in Fig.~\ref{fig:one_line_tn}. It is a variant of the well-known Nordic test system that has been recently revised by the IEEE Power System Dynamic Performance Committee. The system is separated into four zones: \textit{Equiv}, \textit{North}, \textit{Central}, and \textit{South}. Most of the generation is situated in the North and consists of hydropower plants. The rest of the generation, in the \textit{Central} and \textit{South} areas, consists of thermal power plants. Most of the consumption is located in the \textit{Central} and \textit{South} areas. \textit{Equiv} is an equivalent of an external system connected to the North area. In this model, the distribution networks are represented as aggregated loads (see Fig.~\ref{fig:split}, left-hand side). The TN model includes 74 buses: 32 at the transmission level, 20 are generator terminal buses, and 22 buses at the distribution level (medium voltage side of distribution transformers) where the aggregated loads are connected. It includes 102 branches, among which 22 are distribution and 20 step-up transformers. 

\begin{figure}
    \centering
    \includegraphics[width=\columnwidth]{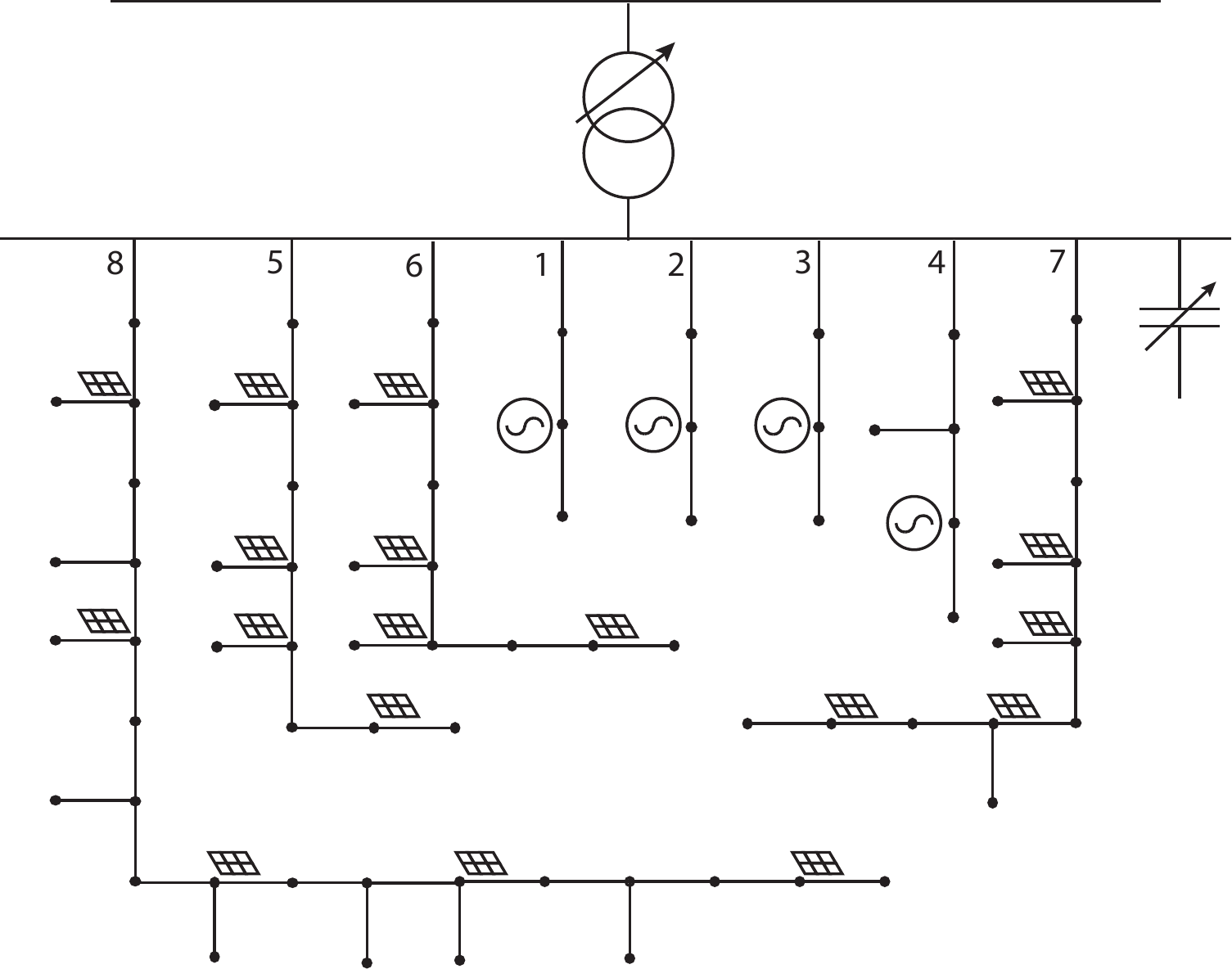}\caption{One-line diagram of the distribution network \cite{UKDGS}}\label{fig:one_line_dn}
\end{figure}

The model used for the DNs was developed by the Centre for Sustainable Electricity and Distributed Generation (SEDG)~\cite{UKDGS}. It represents a radial 11kV urban network fed from a 33kV supply point and its one-line diagram is shown in Fig.~\ref{fig:one_line_dn}. The system has been modified to include two types of DGs and to accommodate higher loading levels. The feeders 1-4 are serving bigger consumers and the DGs connected there are considered to be controllable micro-turbines or small synchronous machines. The remaining feeders are serving residential consumers and the DGs consist of aggregate models of rooftop photovoltaic (PV) systems, thus uncontrolled and generating at their maximum power with unity power factor. Overall, the DN system includes 75 buses with 4 DGs on feeders 1-4 and 18 aggregate PV systems on the remaining feeders.

\section{Key parameters}\label{sec:key_parameters}

In this section, we present some key parameters that are used to generate the final test system. These user-defined parameters are located in the file \textit{parameters.m}, at the root of the toolbox.

\begin{itemize}
    \item \emph{Penetration level}: This value defines the percentage of the active power demand to be generated by the DGs in each DN. It is defined as the ratio between the total active power injected by the DGs and the total power demand of the loads served by the DN, i.e.
    \begin{equation}
        PL = \frac{\sum_{i=1}^{N_{DG}} P_{DGi}}{\sum_{j=1}^{N_{Dj}} P_{D}}
    \end{equation}
    where $N_{DG}$ (resp. $N_{D}$) is the number of DGs (resp. loads) within the DN. It has to be noted that the penetration level is defined \textit{per DN} and not on the entire combined T\&D system. Due to the existence of loads connected at the TN level, the penetration of the combined system will be less than this value.
    
    \item \emph{Generation Split}: This parameter defines the way of splitting the generation between the larger DGs (feeders 1-4 in Fig.~\ref{fig:one_line_dn}) and the smaller ones (feeders 5-8 in Fig.~\ref{fig:one_line_dn}). It allows assigning higher generation levels to the controllable units (feeders 1-4) or to the uncontrollable units (feeders 5-8).
    
    \item \emph{Constant load}: This parameter defines which possible vision of the future power grid is considered for the generated T\&D system. The options are:
    \begin{itemize}
    \item Scenario 1 (\textit{constant load='false'}) assumes that the growth of distributed generation will outpace the growth of electricity demand in DNs. If we consider this from the TN point of view, it means that the net demand of the DNs will decrease. If the penetration level is very high, a reverse power flow can be observed from the DNs to the TN. This option modifies the TN power flows as given in~\cite{PES-TR19}.
    \item Scenario 2 (\textit{constant load='true'}) assumes that the electricity demand will also increase (e.g., because of electric vehicles) alongside distributed generation. Thus, the excess demand will be covered by the DGs and the TN will see the same total demand from the DNs (hence 'constant load'). This option does not modify the TN power flows as defined in~\cite{PES-TR19}.
    \end{itemize}
    
    \item \emph{Random}: Since the same DN template is used to replace all of the TN aggregate loads, this can lead to artificial synchronisations between the various DNs. This setting allows for the parameters of the DNs (e.g., penetration level, generator split, etc.) to be slightly varied around the original values (maximum $\pm5\%$), thus introducing some diversity between DNs.
    
    \item \emph{Large system}: In the default setting (\textit{Large system='true'}), all the loads of the original TN except the ones in the \textit{Equiv} area (on buses 71 and 72 in Fig.~\ref{fig:one_line_tn}) are systematically replaced by detailed DNs. This leads to a T\&D system of approximately 22000 buses. Alternatively, if the parameter is set to \textit{'false'}, then only the loads in the \textit{Central} area are replaced, thus leading to a T\&D system of approximately 15000 buses.
    
    \item \emph{Oversize}: This option allows to oversize the power consumed by each DN, thus creating several security violations. This is useful when the user needs a severely congested test system to test management and operation techniques relying heavily on DGs. As a consequence of the overloading, the number of detailed DNs required to replace each TN load is decreased and consequently also the number of buses in the combined system.
    
    \item \emph{Run OPF}: Determines if the user wants to optimise the operating point of the generated T\&D system before exporting the data. The optimisation procedure is further detailed in \prettyref{sec:OPF}.
    
    \item \emph{Export format}: Defines the format in which the power flow data is exported (MATPOWER or a custom exporter). Two example custom exporters (one for power flow and one for time-domain dynamic simulations) are provided in \prettyref{sec:export}.
    
\end{itemize}

\section{Methodology}\label{sec:Methodology}

In this section, the general methodology of generating the combined T\&D system is presented. As mentioned earlier, \tdnetgen~systematically replaces the aggregated loads of the original TN system with detailed DNs, based on the template DN model presented in \prettyref{sec:systems}. The procedure is summarised here and detailed below:

\begin{enumerate}[label=\Alph*)]
\item Initialize the TN model. Compute the amount of load, per TN bus, that needs to be replaced by detailed DNs based on the \emph{Large system} parameter.
\item Initialize the template DN system and find the maximum DN capacity while avoiding voltage violations. Using the maximum DN capacity and the load per TN bus, calculate the minimum number of DNs required to replace the aggregated loads. Generate the corresponding DN models and compute their individual capacity.
\item Using the parameters \emph{Penetration level}, \emph{Generation split}, \emph{Constant load}, \emph{Random}, and \emph{Oversize}, and their capacity customize each DN operating point by scaling the load and the DGs.
\item Generate the combined T\&D system by interconnecting the models, treating the naming conventions, and ensuring continuity over the boundaries (distribution transformers).
\item \textbf{(Optional)} If the \emph{Run OPF} parameter is selected, optimise the combined T\&D system operating point to reduce generation cost and alleviate voltage problems.
\item Export the combined system data in MATPOWER format or one of the custom exporters, as selected by \emph{Export format}.
\end{enumerate}

\begin{figure}
\begin{flushleft}
\smartdiagramset{%
  set color list={teal!60, teal!60, teal!60, teal!60, red!60, teal!60},
  module x sep=3.0,
  back arrow disabled=true,
  text width=4.5cm,
  additions={
    additional item offset=0.5cm,
    additional item border color=gray,
    additional item font=\small,
    additional arrow color=teal!60,
    additional item text width=2.5cm,
    additional item bottom color=teal!60,
    additional item shadow=drop shadow,
  }
}
\smartdiagramadd[flow diagram:vertical]{%
  Solve the TN power flow (master), {According to the user parameters, solve the DN power flows separately (slaves)}, Replace the TN aggregated loads by DNs in parallel, Solve power flow for the entire T\&D system, Voltages within limits?, Run OPF or export results%
}{right of module4/Update turns ratio of the OLTCs}
\smartdiagramconnect{->}{additional-module1/module4}
\begin{tikzpicture}[remember picture,overlay]
    \draw[additional item arrow type] (module5.east) |- ([yshift=-11mm]additional-module1.south) -- (additional-module1);
  \end{tikzpicture}
\end{flushleft}
\caption{Master-slave approach for the power flow solution}
\label{fig:master_slave}
\end{figure}
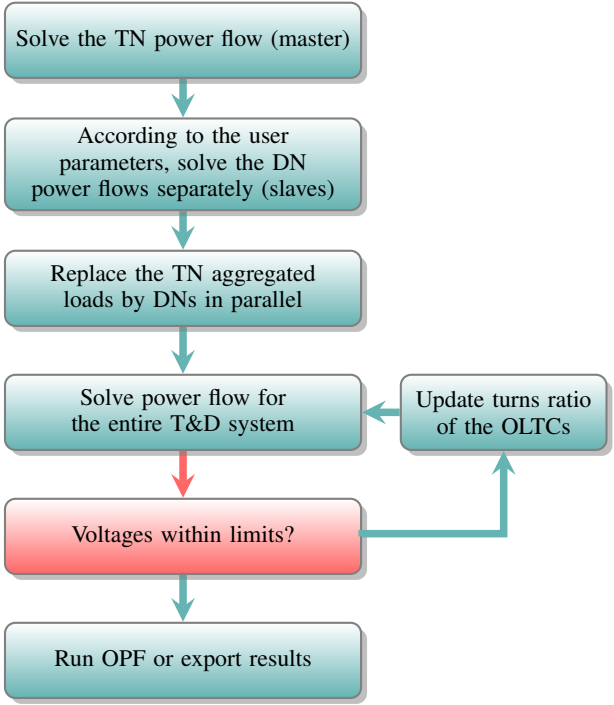

\subsection{Transmission network initialization}

To decrease the complexity of the T\&D model generation, a master-slave approach is followed, illustrated in Fig. \ref{fig:master_slave}. This approach is only used for the construction of the T\&D model. Once the model is exported, the user can analyse it using standard integrated approaches (e.g., MATPOWER's standard power-flow solvers).

First, the TN model (located under \textit{input\_data/tn\_template.mat}) is loaded in MATPOWER and a power-flow solution is performed to compute the voltages at the TN load buses. Then, depending on the \textit{Large system} parameter, it is selected which TN loads will be replaced by detailed DNs. That is, either all the loads in the system will be replaced or only the ones in the Central area.

This approach allows the user to modify the TN operating point before calling \tdnetgen~in order to test different operating conditions. For instance, the parameters of the large TN generators could be modified and the TN load consumption changed to create different scenarios. However, the TN load buses and names should remain unchanged as the following procedure depends on these.

\subsection{Distribution network initialization}

In a second step, the DN template model (located in \textit{input\_data/dn\_template.mat}) is loaded into MATPOWER. The maximum capacity of the DN is calculated by gradually scaling the DN loads under constant power factor while keeping the output of the DGs to zero and making sure that the voltage constraints are not violated. Finally, if the parameter \textit{oversize} is larger than $1.0$, all the DN loads are scaled by this factor, ignoring the voltage violations.

\begin{figure}
    \centering
    \includegraphics[width=\columnwidth]{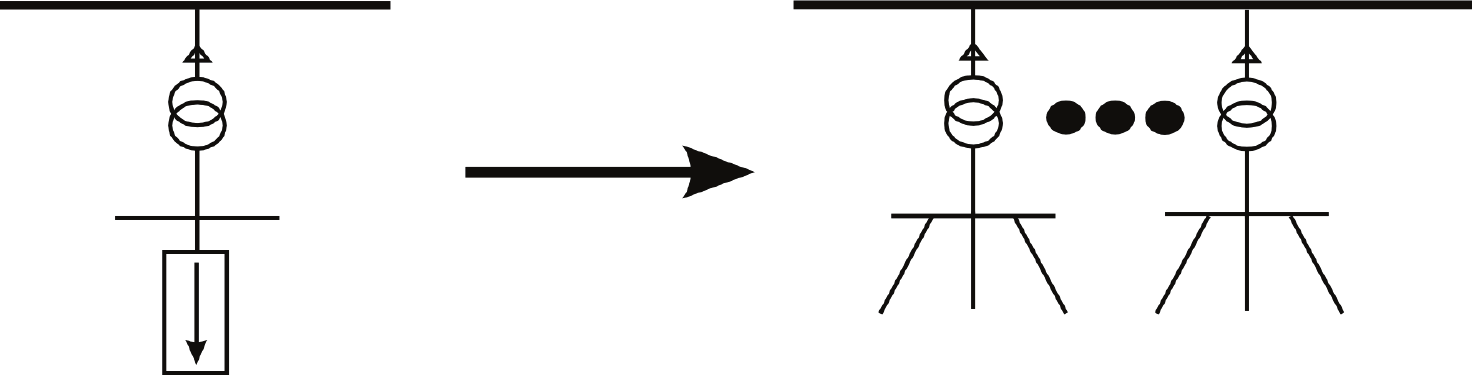}\caption{Aggregated TN loads replaced by detailed DNs connected in parallel on the same TN bus}\label{fig:split}
\end{figure}

Then, the number of DNs needed to replace each aggregated TN load (as shown in Fig.~\ref{fig:split}) is computed by dividing the TN load consumed power with the DN maximum capacity and rounding up to the nearest integer. The implicit assumptions made in this step are that i) the original TN loads are pure loads without any aggregated distributed generation, and ii) the DNs used to replace the TN loads are operating close to their maximum capacity (or oversized). Finally, using the number of DNs per TN bus, the individual DN models are generated and their load consumption is calculated.

Similarly to the TN model, this approach allows the user to modify the DN parameters before executing \tdnetgen. However, much of the scaling and optimisation procedure that follows relies on the exact naming conventions and topology of the DN (location of DGs), thus these should be kept for the modified system.

\subsection{Customization of distribution networks}\label{sec:DN_custom}

In this step, the created DN models are customised according to the parameters defined by the user. The amount of renewable generation is defined by the \textit{penetration level} and the DN consumption; its allocation is determined by the generation splitting parameter, which allows the user to assign more or less power to the controllable DGs in feeders 1-4. Then, if the \textit{Random} flag has been set by the user, a randomization of the DN parameters is performed to make sure that the DN models are not identical. In addition, if the \textit{constant load} parameter is set to true, the active power demand of the DN models is increased to match the introduced DGs, so that the total demand seen by the TN remains the same before and after the replacement of the TN loads by the DN models. At the end of this step, a set of DN models with the user defined characteristics is available to be integrated into the TN.

\subsection{Combined system generation}\label{sec:combined}

As mentioned previously, the integration of the DNs is achieved by replacing the aggregated TN loads by several DNs in parallel (see Fig.~\ref{fig:split}). 

Before connecting the DN models to the TN, the conditions over the boundary elements (distribution transformers) need to be consistent. That is, the magnitude and angle of the high voltage bus to which the DN will be attached (shown in Fig.~\ref{fig:one_line_dn}) needs to match the voltage used for computing the DN power flows in \prettyref{sec:DN_custom}. Consequently, the angles of the other DN buses need to be adjusted accordingly. 

Moreover, the distribution transformers between the TN and the DNs (see Fig.~\ref{fig:one_line_dn}) are considered to be equipped with On-Load Tap-Changing (OLTC) controllers and used to regulate the voltage of the DNs. Basically, the transformer ratio is modified in discrete steps to maintain the voltage at the low-voltage side of the transformer within some deadband as follows for the $i$-th transformer:
\begin{equation}\label{eq:OLTC}
\begin{cases} 
\mbox{if } V^C_i > V^{set}_i+\frac{DB_i}{2} & r_i=r_i+1 \\ 
\mbox{if } V^C_i < V^{set}_i-\frac{DB_i}{2} & r_i=r_i-1
\end{cases}
\end{equation}
where $V^{set}_i$ and $DB_i$ are the controller setpoint and deadband, respectively; $V^C_i$ is the controlled voltage; and, $r_i$ is the discrete tap defining the transformer ratio.

After the customization of the DNs in \prettyref{sec:DN_custom}, the DN voltage controlled by the OLTC ($V^C_i$) might be out of the deadband and require adjusting the transformer ratio. Unfortunately, MATPOWER does not support OLTC transformers. Thus, their functionality was implemented externally with an iterative procedure: a combined T\&D power flow solution is performed, then the voltages of the DN substations are checked and the OLTC ratios adjusted accordingly. This is followed by another combined system power flow solution to update the values. This sequence is repeated until the DN voltage set-points (which can be changed by the user in \textit{parameters.m}) are reached. Some safeguards in the form of a maximum allowed changes have been implemented to avoid infinite cycling of the OLTCs.

Finally, the combined T\&D network model is generated. In the case of \textit{constant load} set to true, the TN power flows and voltages will be unchanged. Alternatively, the TN power flows will be different and at high penetration levels even reversed. Moreover, due to the integration of the DGs, there might be some voltage violations both in the DNs as well as in the TN. The latter only if \textit{constant load} is set to false.

\subsection{Operating point optimization}\label{sec:OPF}

The resulting T\&D system from the previously described procedure might lead to increased generation costs as well as voltage violations. \tdnetgen~provides the option to run an AC OPF to optimise the combined system operation and alleviate any voltage problems before exporting the data. Unlike the previous steps, this is an optional step (controlled by the \emph{Run OPF} parameter) as the users might want to implement their own OPF algorithm or test some operational schemes for alleviating voltage problems.

To optimise the operating point, the following OPF problem needs to be solved:
\begin{mini!}[2]
  {\bm x,\bm c,\bm r}{f(\bm x,\bm c,\bm r)\label{eq:OPF-obj}}{\label{eq:OPF-opti}}{}
  \addConstraint{\bm g(\bm x,\bm c,\bm r)=\bm 0\label{eq:OPF-Eq}}{}
  \addConstraint{\bm h(\bm x,\bm c,\bm r)\leq\bm 0\label{eq:OPF-Ineq}}{}
  \addConstraint{\underline{\bm c}\leq\bm c \leq \overline{\bm c}\label{eq:OPF-Ineq2}}{}
  \addConstraint{\bm r=[r_1 \ldots r_i \ldots r_{nd}]\label{eq:OPF-disc}}{}
  \addConstraint{r_i \in \{r_i^1 \ldots r_i^j \ldots r_i^{max}\},\,\, \forall i=1,\ldots,r_{nd}\label{eq:OPF-disc2}}{}
\end{mini!}
where $\bm x$ is the vector of state variables (i.e., voltage magnitudes and phases at all buses), $\bm c$ is an $n_c$ dimensional vector of continuous control variables (i.e., active and reactive powers of generators) and $\underline{\bm c}$ (resp. $\overline{\bm c}$) is its corresponding vector of lower (resp. upper) bounds, $r$ is an $n_d$ dimensional vector of discrete control variables (i.e., the OLTC transformer ratios), $r_i^j$ is the $j$-th discrete value of discrete variable $r_i$, $r_i^{max}$ is the number of discrete positions of the OLTC, $\bm f(\cdot)$ is the objective function, $\bm g(\cdot)$ and $\bm h(\cdot)$ are vectors of functions which model equality and inequality constraints. 

\begin{figure}
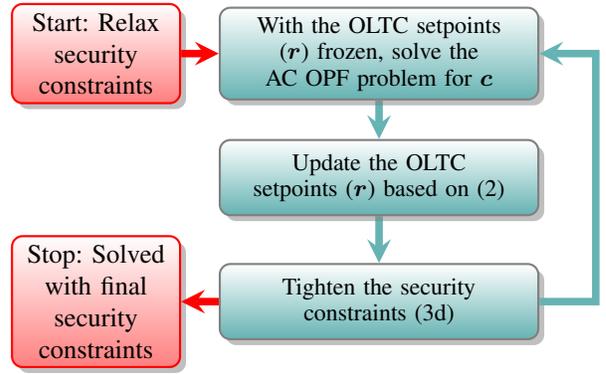

\begin{flushright}
\smartdiagramset{%
  uniform color list=teal!60 for 3 items,
  module x sep=3.0,
  back arrow distance=0.75,
  text width=4cm,
  additions={
    additional item offset=0.5cm,
    additional item border color=red,
    additional arrow color=red,
    additional item text width=2cm,
    additional item bottom color=red!50,
    additional item shadow=drop shadow,
  }
}
\smartdiagramadd[flow diagram:vertical]{%
  {With the OLTC setpoints ($\bm r$) frozen, solve the AC OPF problem for $\bm c$}, Update the OLTC setpoints ($\bm r$) based on \eqref{eq:OLTC}, Tighten the security constraints \eqref{eq:OPF-Ineq2}%
}{left of module1/Start: Relax security constraints,%
left of module3/Stop: Solved with final security constraints}
\smartdiagramconnect{<-}{additional-module2/module3}
\smartdiagramconnect{<-}{module1/additional-module1}
\end{flushright}
\caption{Iterative relaxation technique for the OPF solution}\label{fig:OPF-SOl}
\end{figure}

The objective function \eqref{eq:OPF-obj} is to minimise the cost of supplying the load. Only the large TN generators and the DGs located in feeders 1-4 of each DN are dispatched. That is, the PV systems in feeders 5-8 do not participate and are assumed to always operate at maximum power and unity power factor. The generator quadratic cost functions were taken from \cite{Gao2015} and \cite{Feng2014}, for the TN and DN generators respectively. Their values are defined in \textit{functions/add\_gen\_costs\_for\_OPF.m} and can be modified by the user.

The equality constraints \eqref{eq:OPF-Eq} are the AC bus power flow equations, the inequality constraints \eqref{eq:OPF-Ineq} refer to operational limits (i.e., the voltage magnitudes security constraints), the inequality constraints \eqref{eq:OPF-Ineq2} refer to physical limits of equipment (i.e., bounds on generators active/reactive powers), and the constraints \eqref{eq:OPF-disc} describe the discrete variable values of the OLTC controllers.

Efficiently solving a large-scale, mixed-integer, optimisation problem as described by \eqref{eq:OPF-opti} can be challenging. Several methods have been proposed in the literature based on the round-off strategy, using penalty terms, or several other heuristic methods (several examples can be found in \cite{Capitanescu2010,Platbrood2014} and their references). However, the focus of this paper is on providing a parametrizable, large-scale, test-system and not on developing solution techniques for OPF problems with discrete decision variables. Thus, a simple, heuristic, iterative relaxation method is employed in this work.

The procedure used in \tdnetgen~is summarised in Fig.~\ref{fig:OPF-SOl}. First, the voltage constraints \eqref{eq:OPF-Ineq} are relaxed and the discrete variables \eqref{eq:OPF-disc} are fixed constant. The latter transforms the mixed-integer OPF into a "standard AC OPF problem" that  is solved with the included MATPOWER algorithm to compute an estimate of the continuous control variables $\bm c$. Based on this estimate, the discrete variables $\bm r$ are updated using the control rules of \eqref{eq:OLTC}. Finally, the security constraints are tightened and the procedure is repeated until the problem is solved with the final constraints, selected by the user, and the OLTC control rules \eqref{eq:OLTC} are satisfied.

Starting from MATPOWER version 6.0, a new tool called MOST~\cite{Sanchez2013} is also provided for optimal scheduling including uncertainty in demand and RES generation. However, this has not been tested in this work.

\subsection{Data exporters}\label{sec:export}

The final step is to export the combined system into the \textit{output\_data} folder. In the default format, \tdnetgen~exports a MATPOWER case file that can be loaded, modified, and analysed, without the need of the \tdnetgen~being present.

In addition, the toolbox allows the user to build a custom exporter to any format, depending on the simulation software used. \tdnetgen~provides two examples of custom exporters, provided in the \textit{custom\_data} folder. The first is the power flow program ARTERE developed at the University of Liege and available at \cite{ARTERE}. Compared to MATPOWER, this software includes the modelling and treatment of OLTC transformers and is computationally more efficient. 

The second custom exporter is for the academic time-domain, dynamic simulation software RAMSES~\cite{Aristidou2015f}. In order to execute a dynamic simulation, the dynamic data for both the TN generators and controllers as well as the DGs, are required. For the TN, the dynamic data are taken from~\cite{PES-TR19}. For the DNs, the DGs in feeders 1-4 are modelled as small synchronous machines following~\cite{VVC13a}, while the DGs in feeders 5-8 use the distributed PV system model (PVD)~\cite{Elliott2015}. The initialization of the dynamic models takes place in RAMSES based on the operating point provided by \tdnetgen.

Using the above two examples, users can create their own exporters to a variety of software packages. One of the future developments will be the implementation of an exporter to CIM format~\cite{CIMPrimer}, to facilitate the integration of the models to other software.

\section{Example Test Cases}\label{sec:Power-Flow-Test-Cases}

In this section, we present some combined T\&D test systems generated by \tdnetgen~focusing on producing some operationally challenging test cases.

\subsection{Power-flow scenarios}

\begin{figure}
    \centering
    \includegraphics[trim=30 10 40 30, clip, width=1\columnwidth]{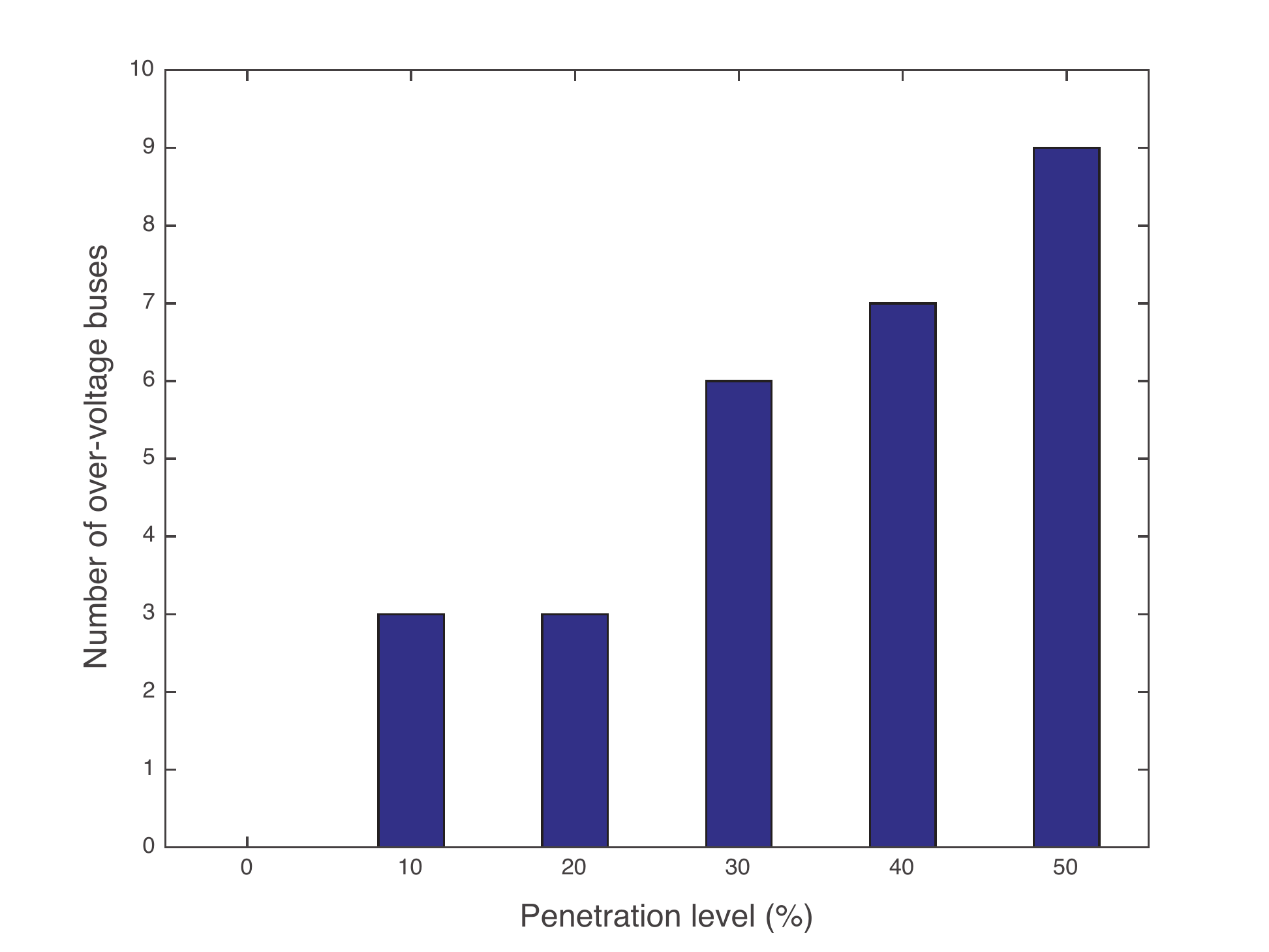}\caption{Over-voltages above 1.15~pu in the TN for different \textit{Penetration level} values [\textit{Constant load='false'}, \textit{Large system='true'}, \textit{Run OPF='false'}]}\label{fig:over}
\end{figure}

One of the main problems associated with increased penetration of DGs and in particular RES are possible over-voltage problems arising at high generation levels. \tdnetgen~is able to generate such scenarios by increasing the penetration level. As the penetration level increases, more buses experience over-voltages, in the DNs but also in the TN. Figure~\ref{fig:over}, shows the number of TN buses with over-voltages (above 1.15 pu) as the penetration level increases. In these scenarios, the operating point produced by \tdnetgen~is not optimised (\textit{Run OPF='false'}). If an OPF solution is selected, then the operating point can be corrected by re-dispatching the TN generators and the DGs in feeders 1-4 of each DN.

\begin{figure}
    \centering
    \includegraphics[trim=30 10 60 30, clip, width=1\columnwidth]{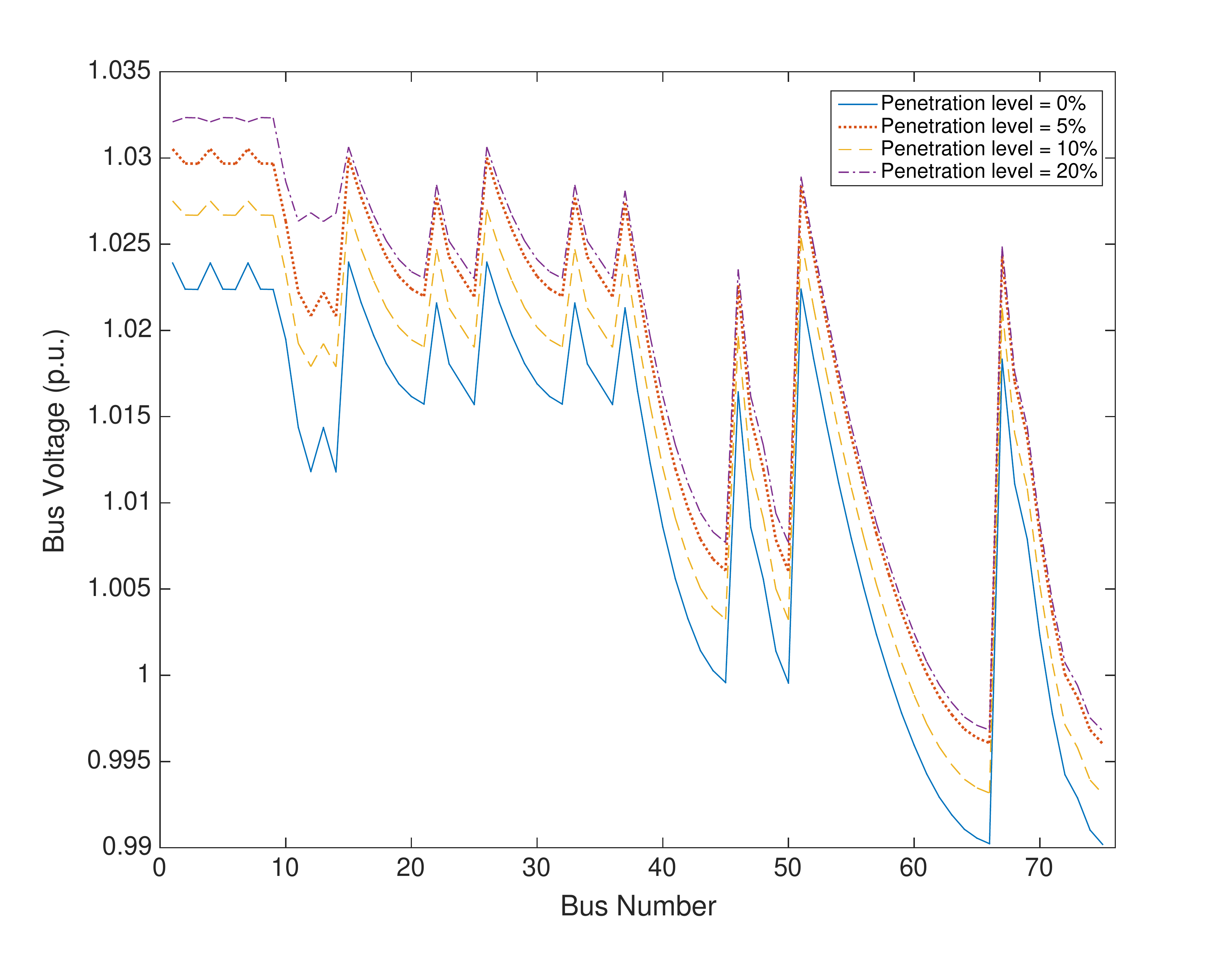}\caption{Voltage profile of DN for low penetration levels [\textit{Constant load='false'}, \textit{Large system='true'}, \textit{Run OPF='false'}]}\label{fig:pl0_20}
\end{figure}

\begin{figure}
    \centering
    \includegraphics[trim=30 10 60 30, clip, width=1\columnwidth]{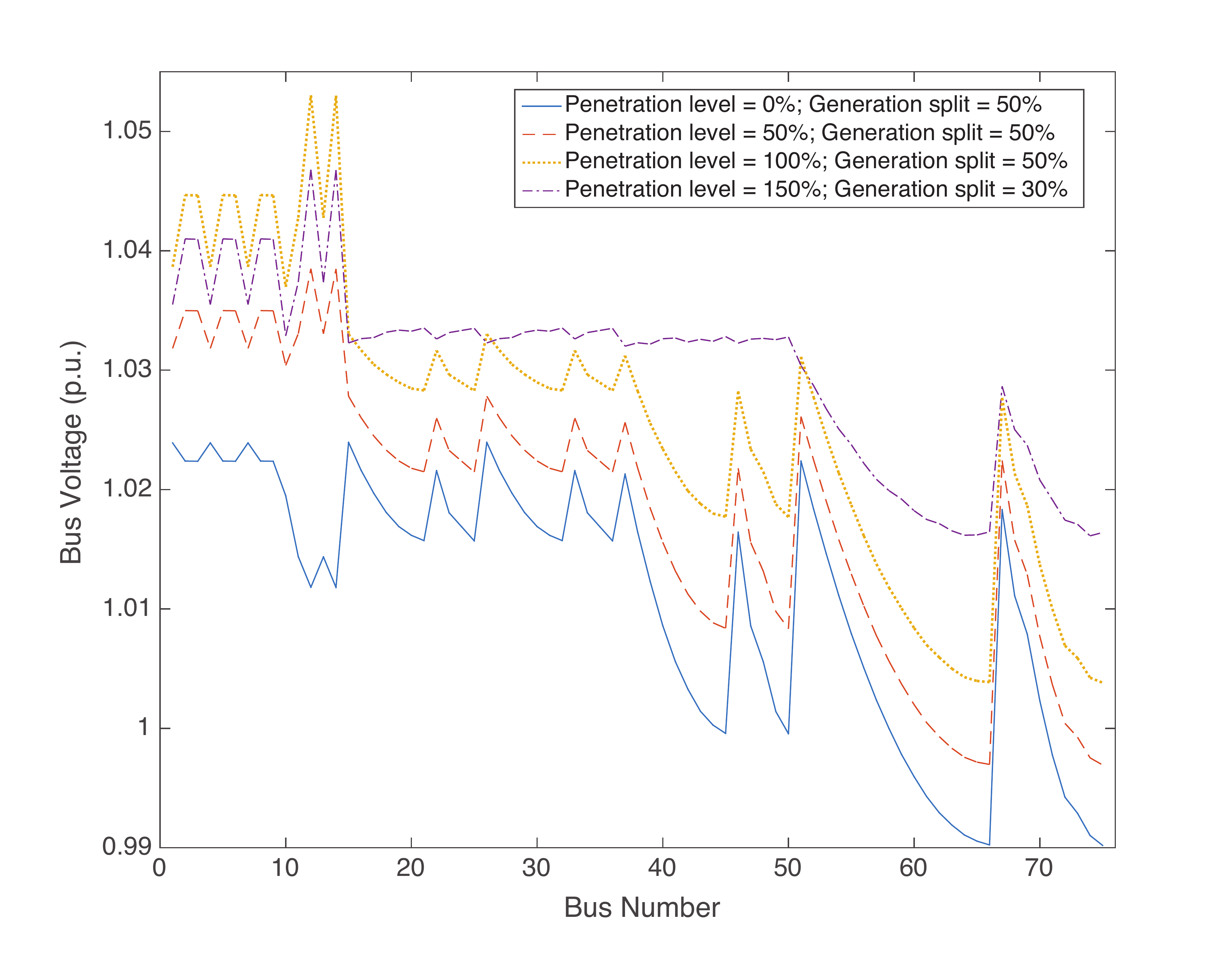}\caption{Voltage profile of DN for high penetration levels [\textit{Constant load='false'}, \textit{Large system='true'}, \textit{Run OPF='false'}]}\label{fig:pl25_75}
\end{figure}

Figures~\ref{fig:pl0_20} and \ref{fig:pl25_75} show the voltage profiles at all the buses located in a specific DN for different penetration levels. In these figures, the 75 DN buses are numbered consecutively starting from feeder 1 up to 8 and from the bus closest to the DN substation to the one furthest (see Fig.~\ref{fig:one_line_dn}). The continuous curve represents the voltage profile of the DN without any DG and where the voltage set-point of the OLTC is at 1.03~pu\footnote{There is a slight variation of the actual voltage controlled by the OLTC over the different scenarios due to the effect of the deadband.}. It can be seen that as the penetration level increases, the voltage profile of the feeder is changing.

\begin{figure}
    \centering
    \includegraphics[trim=30 80 30 10, clip, width=1\columnwidth]{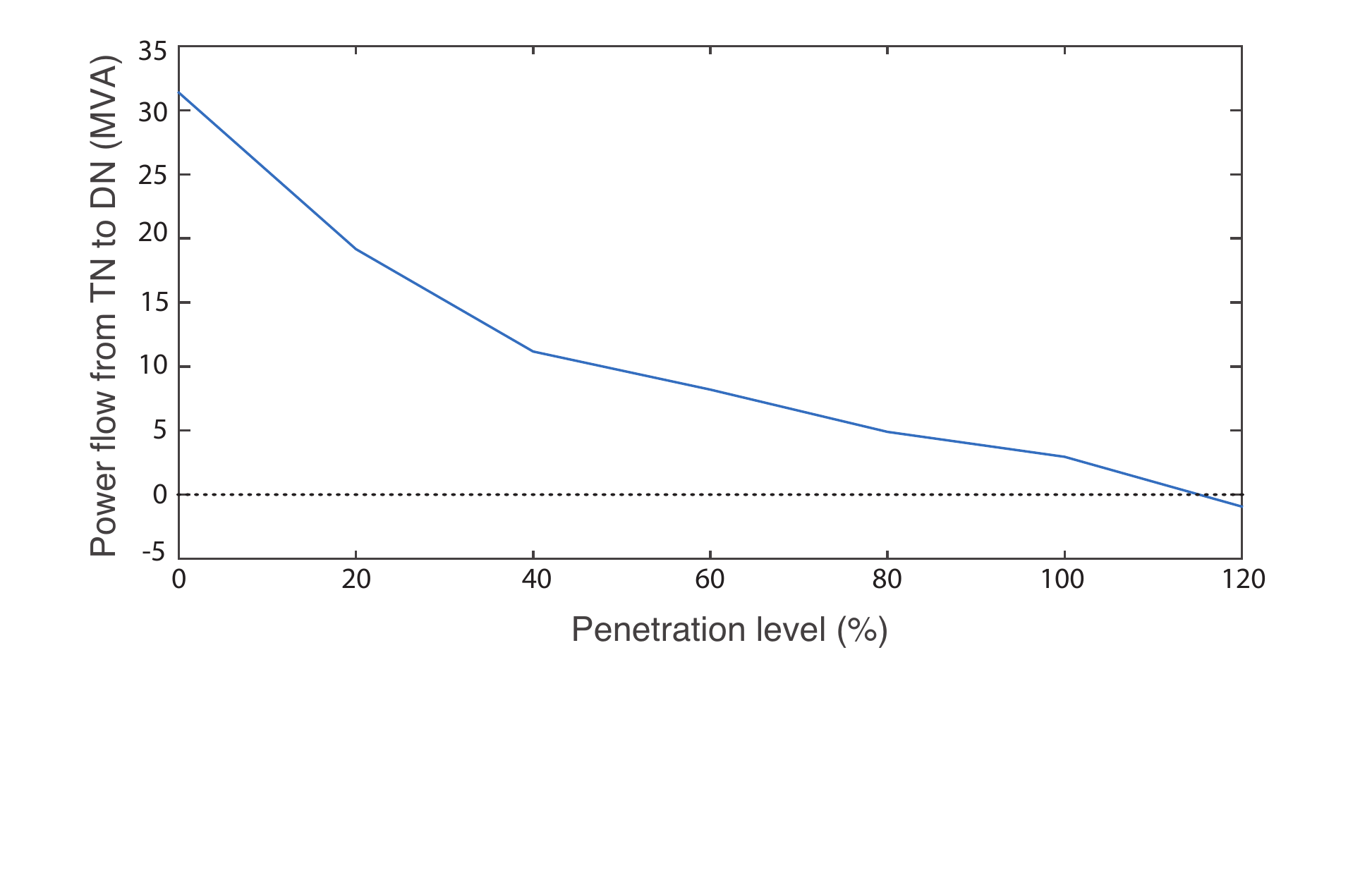}\caption{Power flow from the transmission to the distribution network [\textit{Constant load='false'}, \textit{Large system='true'}, \textit{Run OPF='false'}]}\label{fig:rev_pf}
\end{figure}

As the penetration level increases, the total power demand of the DNs from the TN decreases. In cases of higher than 100\% penetration, when the DGs cover the local load consumption as well as the DN losses, then a reverse power flow arises. In these extreme cases, the power flows from the DNs to the TN and it can cause voltage violations and challenge the effectiveness of existing management, control, and protection schemes. Figure~\ref{fig:rev_pf} shows the total power transferred from the TN to all the DNs for different penetration levels. It can be seen that at about 115\%, the DNs cover their own local consumption and losses and there is no power exchange with the TN. It should be noted that individual DNs might reach a reverse power flow condition at lower or higher penetration levels due to their different operating points.

\subsection{A dynamic simulation scenario using RAMSES}

As explained in \prettyref{sec:export}, custom exporters can be used to translate the data into other formats used in other software than MATPOWER. One of the example exporters with the toolbox produces the data to be used in RAMSES, an academic, time-domain, dynamic simulation software~\cite{Aristidou2015f}. This simulator uses a topologically-based, domain decomposition method to partition the system into the TN and the multiple DNs. Then, it solves the sub-problems defined over each sub-domain at each discrete time instance, while treating the interface variables with a Schur-complement approach. Parallel computing techniques are employed to accelerate the system simulation.

A T\&D model was generated with the parameters \textit{Constant load='true'}, \textit{Large system='false'}, \textit{Run OPF='false'}, \textit{Penetration level=50\%}, \textit{Oversize=2.0}. The data was exported into RAMSES format, leading to a system with 141 DNs in the Central area, with a total of 10782 buses, 20 large (connected to the TN) and 564 small synchronous generators (connected to the DN feeders 1-4), 2538 distributed PV systems (connected to the DN feeders 5-8), and 10595 voltage-dependent loads. The considered disturbance is a 5-cycle (100~ms) short-circuit near bus 4032, cleared by opening line 4032-4042. The response is simulated over a horizon of 250~s with a time-step of 1 cycle (20~ms).

\begin{figure}
    \centering
    \includegraphics[width=\columnwidth]{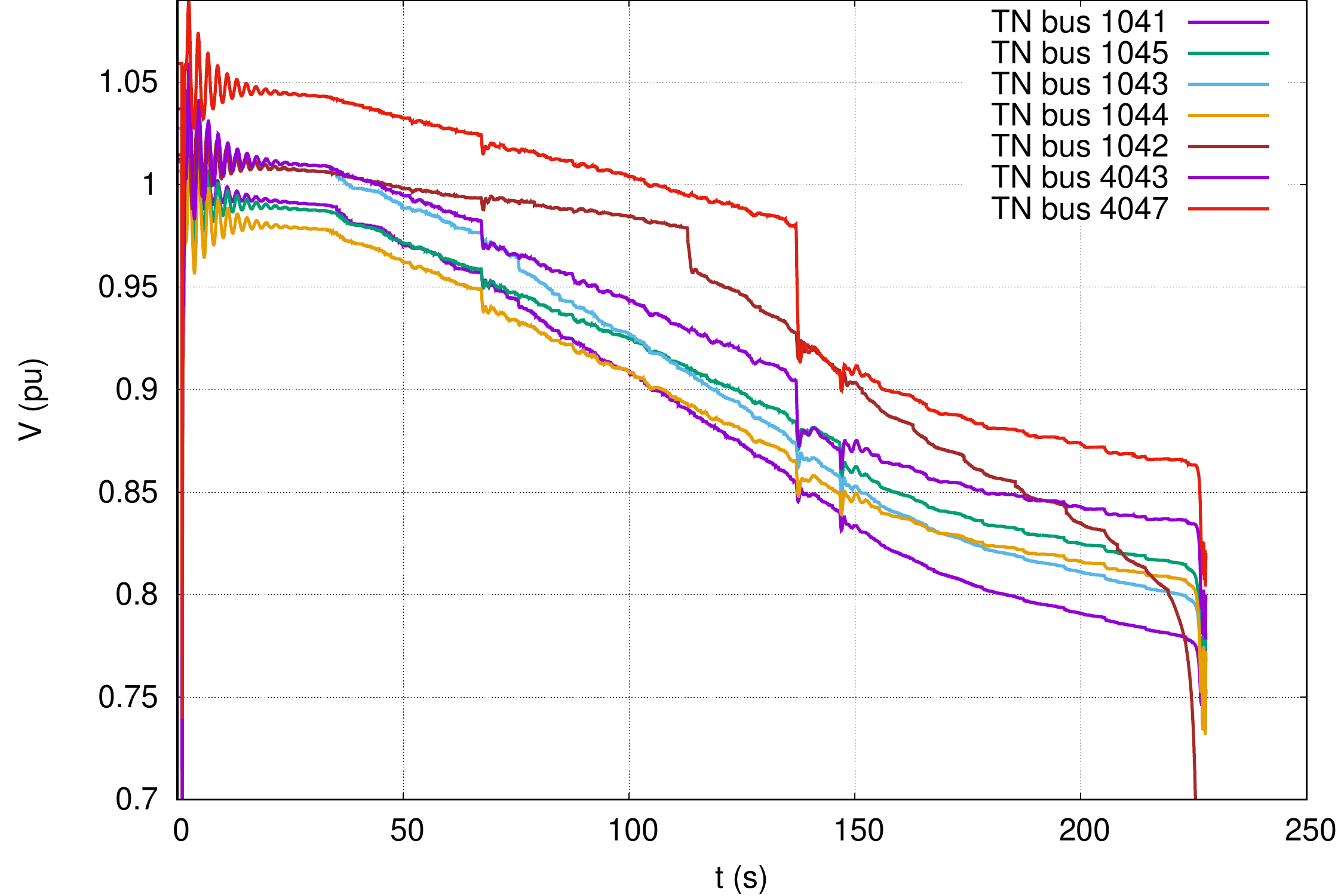}\caption{Voltages at various TN buses during the dynamic simulation}\label{fig:ramsesTN}
\end{figure}

\begin{figure}
    \centering
    \includegraphics[width=\columnwidth]{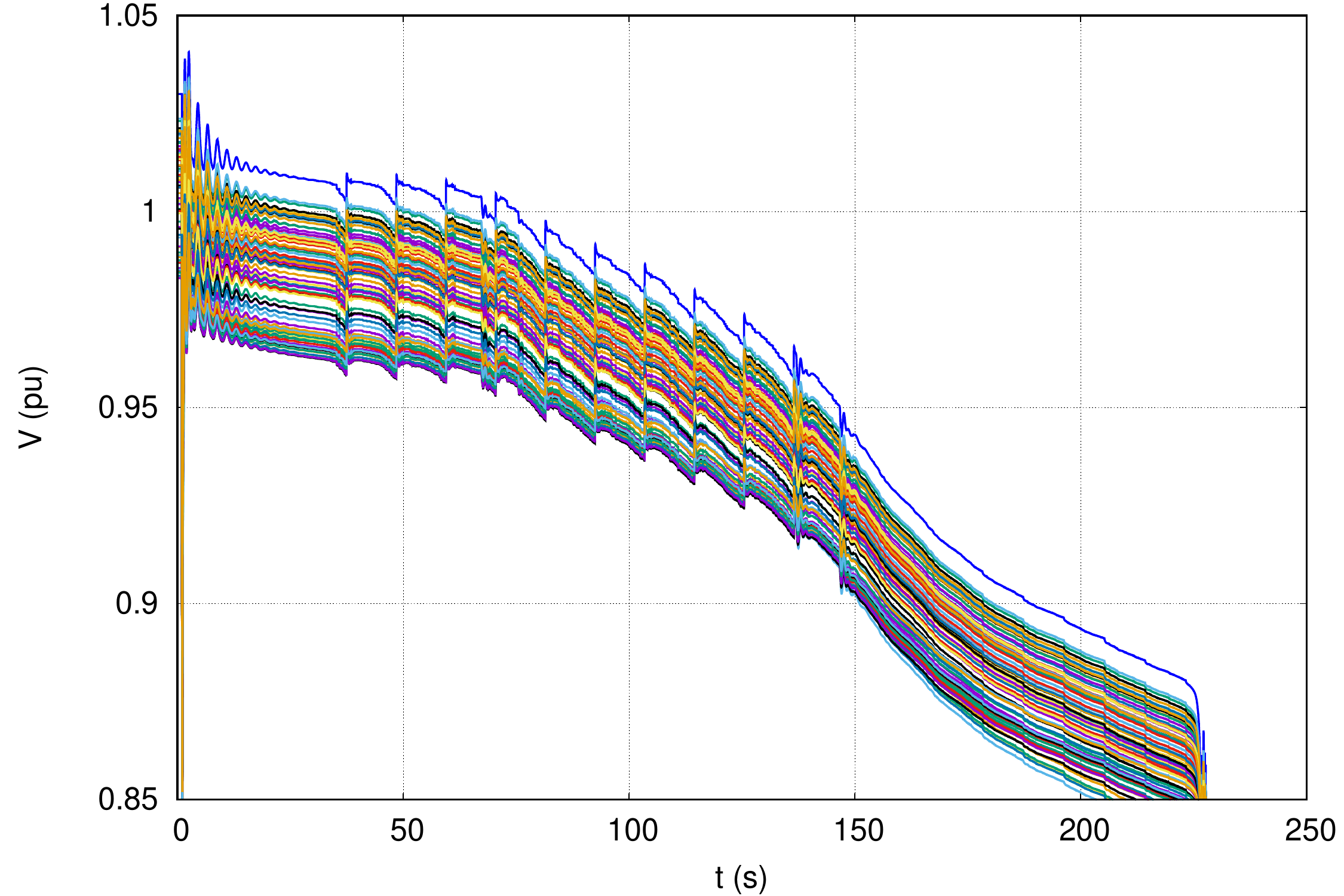}\caption{Voltages at various DN buses within the same DN during the dynamic simulation}\label{fig:ramsesDN}
\end{figure}

This scenario leads to a long-term voltage collapse driven by the load-power restoration caused by the OLTCs restoring the DN voltages. Figure~\ref{fig:ramsesTN} shows the evolution of the TN voltages while Fig.~\ref{fig:ramsesDN} shows the DN voltages at the buses in a DN attached to the TN bus 1041. More information on the analysis, instability detection and corrective control of such unstable scenarios in combined T\&D systems is given in \cite{Aristidou2015d}.

\subsection{Computational cost of \tdnetgen}
Depending on the choice of parameters, the run-time to generate the T\&D system can vary significantly. However, this is a cost occurred only once to generate the test system, which can then be used independently of \tdnetgen. The most computationally intensive task is the OPF solution of the system to optimise the operating point (if selected). Table~\ref{tab:cpu} shows the average execution times to generate a T\&D system\footnote{Acquired on a MacBook Pro laptop with 2.2GHz Intel Core i7, 16GB of RAM, using MATLAB 2015a and MATPOWER 6.0} as well as the average number of times the OLTC setpoints are updated during an execution. The overall execution time does not exceed 4 minutes, using a standard laptop computer.

\begin{table}
\caption{Average computational cost of different \tdnetgen  ~operations}\label{tab:cpu}
\begin{center}
  \begin{tabular}{l l l}
    \hline
    Operation & Run-time & OLTC updates\\
    \hline
    Power Flow & 30 seconds & 22\\
    Optimal Power Flow & 200 seconds & 3\\
    Export data & 15 seconds & -\\
    \hline
  \end{tabular}
\end{center}
\end{table}

\section{Conclusion}\label{sec:conclusion}

In this paper, an open-source MATLAB toolbox was presented, that is able to generate synthetic large-scale T\&D network test cases. These models can be used to develop or test solutions targeting problems specific to combined T\&D systems (optimisation, market clearing, system security, ancillary services by DNs, etc.). The toolbox is based on the widely used MATPOWER software and uses the well-known Nordic system as the TN upon which the combined T\&D system is built. \tdnetgen~is highly parametrizable, allowing to generate models with a variety of characteristics, replicating common problems in systems with high penetration of RES. In addition, medium- ($\sim 10000$ buses) to large-scale ($\sim 22000$ buses) test-system models can be produced, allowing to test the scalability of algorithms proposed by the users. Moreover, the model data can be shared among researchers to provide a common testing platform.

Finally, the open-source nature of the toolbox allows the users to expand the code, adding new functionality, or building custom export functions to import the generated models into their own software. The latest version of the toolbox can be found at \cite{TDNETGEN}.

\section*{Acknowledgment}
\addcontentsline{toc}{section}{Acknowledgment}

The authors would like to thank Prof Thierry Van Cutsem for the valuable input in the development of the toolbox and Gilles Chaspierre for being the first user of the toolbox and identifying several bugs.

\bibliographystyle{IEEEtran}
\bibliography{Mendeley_group}

\newpage

\begin{IEEEbiography}%
[{\includegraphics[width=1in,height=1.25in,clip,keepaspectratio]{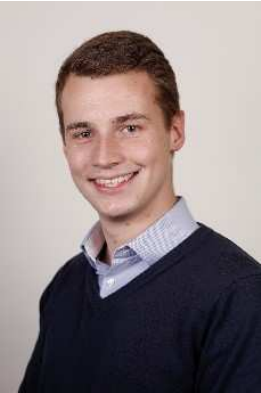}}]
 {Nicolas Pilatte} was born in Etterbeek, Belgium. He received his B.Sc. in Electrical Engineering from the University of Mons, Belgium in 2015. He is currently enrolled as a M.Sc. student in Energy Science and Technology at the ETH Zurich, Switzerland. His research interests include power system stability and integration of renewable technologies.
\end{IEEEbiography}
\begin{IEEEbiography}
    [{\includegraphics[width=1in,height=1.25in,clip,keepaspectratio]{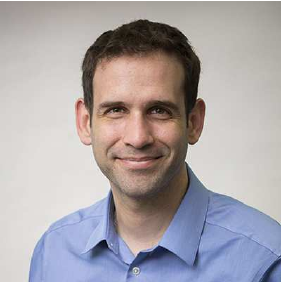}}]{Petros Aristidou}(M\textquoteright{}10) is a Lecturer at the University of Leeds, UK. Before that he was a postdoctoral researcher at ETH Zurich, Switzerland. He holds a PhD from the University of Li\`ege, Belgium (2015) and a Diploma in Electrical and Computer Engineering from the National Technical University of Athens (NTUA), Greece (2010). His research interests include power system dynamics, control, and simulation.
\end{IEEEbiography}
\begin{IEEEbiography}%
[{\includegraphics[width=1in,height=1.25in,clip,keepaspectratio]{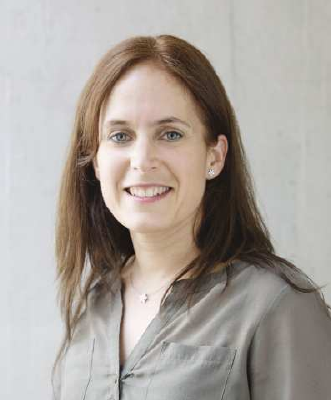}}]
 {Gabriela Hug} (S\textquoteright{}05, M\textquoteright{}08, SM\textquoteright{}14) was born in Baden, Switzerland. She received the M.Sc. degree in electrical engineering in 2004 and the Ph.D. degree in 2008, both from Swiss Federal Institute of Technology (ETH), Zurich, Switzerland. After the Ph.D. degree, she worked in the Special Studies Group of Hydro One, Toronto, ON, Canada, and from 2009 to 2015, she was an Assistant Professor at Carnegie Mellon University, Pittsburgh, PA, USA. She is currently an Associate Professor in the Power Systems Laboratory, ETH Zurich. Her research is dedicated to control and optimization of electric power systems.
\end{IEEEbiography}
\vfill
\end{document}